\documentclass[conference]{IEEEtran}
\IEEEoverridecommandlockouts

\usepackage{cite}
\usepackage{amsmath,amssymb,amsfonts}
\usepackage{algorithmic}
\usepackage{graphicx}
\usepackage{graphics}       
\usepackage{cleveref}
\usepackage{epsfig}
\usepackage{tabulary}
\usepackage{array}
\usepackage{textcomp}
\usepackage{xcolor}
\usepackage{lipsum}
\usepackage{mathtools, cuted}

\begin{document}

\title{Comparison of linear observation techniques for robust load torque estimation in actuators}

\author{Michael Ruderman, Elia Brescia, Luigi P. Savastio, Paolo R. Massenio, David Naso, Giuseppe L. Cascella   
\thanks{M. Ruderman is with Dep. of Engineering Sciences, University of Agder, Norway. He is on annual sabbatical at Polytechnic University of Bari.\newline
Correspondence to: email {\tt\small michael.ruderman@uia.no}}
\thanks{E. Brescia, L. Savastio, P. Massenio, D. Naso, G. Cascella 
are with Dep. of Electrical and Information Engineering, Polytechnic University of Bari, Italy.
}
\thanks{The first author acknowledges the financial support by NEST (Network for Energy Sustainable Transition) foundation.}
}

\maketitle

\bstctlcite{references:BSTcontrol}

\begin{abstract}
The paper addresses the problem of estimating robustly the external load torque in rotary actuator systems, when only the generated motor drive torque and angular displacement are the available input and output. We compare, theoretically and experimentally, two sufficiently established linear observation techniques (i) reduced-order Luenberger observer and (ii) disturbance observer, both using the same identified model of a permanent magnet synchronous motor (PMSM)-based actuator. Our goal is to highlight several aspects related to the implementation, relative degree of the input-torque to estimated-load-torque transfer characteristics, observer open-loop transfer function, and the associated sensitivity (respectively stability margins) with respect to inherently uncertain system plants. Apart from the developed analysis, a detailed experimental case study is demonstrated where the load torque sensor provides reference measurements and allows for evaluation of both observers.
\end{abstract}


\section{Introduction}  
\label{sec:1}

The load torque, or mechanical load in the generalized force coordinates at large, is an essential part of the actuator dynamics and often required to be known in advanced motion control systems \cite{ruderman2020}. A direct use of the load or force sensors between the controlled actuator and an external load is rather an uncommon practice, due to the associated hardware costs and the often cumbersome interfacing for sensor mounting. Therefore, for many years, the question of a robust estimation of the load forces and torques was among the most relevant for a regulated operation of the industrial machinery, robots, and motion control systems as such, cf. e.g. \cite{ohnishi1994}.

In the following, we are focusing on the unknown load torques in rotary motor-driven actuators and address two well established linear observation techniques in a comparative, and hereby theoretical and experimental, way. Our focus is purposefully dedicated to the linear techniques as most tractable and reliable for many applications. Some minimal and pivotal knowledge of the motion dynamics, accessibility of analysis in frequency domain, and comparability of such approaches with regard to the only measured output displacement quantity (affected by the sensor noise) are the main selection criteria. Therefore, neither parallels nor comparisons to nonlinear estimation techniques are made here, although those certainly deserve mention in other studies. The intuitively first choice for a linear (asymptotic) state observer is the Luenberger one, see \cite{luenberger1971} for basics, while a reduced-order version allows to decrease the observer's order, thus facilitating the design and simplifying the overall coupled dynamics of the estimation state variables. For more recent analysis of the Luenberger observer and its nonlinear extension (i.e. Kazantzis–Kravaris–Luenberger observer) we also refer to e.g. \cite{bernard2018}. Since the load torque constitutes a matched (dynamic) perturbation state, another linear estimation technique also appears to be directly a candidate. This one is the disturbance observer \cite{ohishi1987} approach (often denoted as DOB), which is also used as a tool to estimate the accumulated and matched, also sometimes called ``equivalent'', disturbance. For some more recent overview on DOB, with the emphasized transfer and filtering characteristics relevant for analysis, we also refer to e.g. \cite{oboe2018}, while an analysis of the almost necessary and sufficient conditions for robust stability of the closed-loop systems with DOB can be found in \cite{shim2009}. Another potential candidate, which can be called the detection- and isolation-based method \cite{depersis2002}, using the generalized momenta of the equations of motion, reveals also the linear error dynamics, but does not allow directly to decouple nonlinearities and to perform such an analysis as pursued in this work. Worth mentioning, however, is that this applied method was developed and used in \cite{deluca2003} for detecting and isolating the actuator faults, and later in \cite{ruderman2014} for estimating the torsional torque in elastic robot joints.
 
One needs to emphasize that a matched ``equivalent disturbance'', independently of the estimation technique behind, bears naturally the signature of also other i.e. unaccounted matched perturbations, equally as of the nonlinear friction that is essential for almost all actuator systems, see \cite{ruderman2023analysis} for details. However, our current focus remains primarily on estimating the load torque itself rather than decomposing various matched disturbances or estimation-based compensation with a feedback loop. These lines of research can be considered in the future works. To mention is also that some comparative statements about the DOB-based disturbance compensation versus the PID-based one, can be found e.g. in \cite{oboe2018}. As for an integral feedback action, its inability to compensate for e.g. nonlinear friction was recently studied in detail in \cite{ruderman2025}. And for a more generic discussion of the impact of disturbances (including the load and friction forces and torques) in context of the motion control we also refer to \cite{ohnishi1994,ruderman2020}.   
  
The rest of the paper is organized as follows. In Section \ref{sec:2}, we make the modelling assumptions of the actuator dynamics and provide a detailed description of the experimental drive  system used in this work. Section \ref{sec:3} introduces in a summarizing form both linear observation techniques in use and provides all necessary steps for designing the corresponding observers. An analysis of observer in the loop is developed in Section \ref{sec:4}, while highlighting one of the most relevant aspects -- the sensitivity (correspondingly stability related issues) of the open-loop  transfer characteristics of observer in case of the inherent system uncertainties. The experimental evaluation of both estimation techniques with the reference measurement of the load torque, done by means of a dedicated torquemeter, is reported in Section \ref{sec:5}. Conclusions are drawn in Section \ref{sec:6}.

\section{Actuator drive}  
\label{sec:2}

\subsection{Motion dynamics}  
\label{sec:2:sub:1}

We assume the most generic model of a surface-mounted permanent magnet synchronous motor (PMSM)-driven actuator, which is simple as possible and, at the same time, reliable by capturing the main properties of the motion dynamics. The controllable system input is the electromagnetic PMSM drive torque $u(t)$, while the single measurable output available for both -- torque estimation and (generally) feedback control purposes -- is the angular displacement of output shaft $q(t)$. Note that the latter is usually sensed by some optical or magnetic encoder and is naturally subject to a sensor noise. The dynamic behavior of the actuator system is given by
\begin{equation}\label{eq:2:1}
J \ddot{q}(t) + b \dot{q}(t) - \tau(t) = u(t) - C_f \textrm{sign}\bigl( \dot{q}(t) \bigr),
\end{equation}
where the lumped system parameters, the inertia $J$, the linear damping $b$, and the Coulomb friction coefficient $C_f$, are assumed to be sufficiently known from the system identification and/or available technical data. The matched load torque $\tau(t)$ is not measured and, correspondingly, of our prime interest to be estimated. Note that the Coulomb friction term is purposefully regarded on the right-hand-side of \eqref{eq:2:1}, since this is the nonlinear one and it will be decoupled in a model-based way when considering the linear observers. Further we note that the PMSM torque is available from the measured and controlled motor $q$-axis current, i.e. $u(t)=K_m i_q(t)$, where the motor torque constant $K_m$ is assumed to be accurately known and state- correspondingly time-independent. Also no significant torsional elasticities or gear dynamics are assumed for a basic class of the actuator drives \eqref{eq:2:1}.

\subsection{Experimental system}  
\label{sec:2:sub:2}

\begin{figure}[!h]
	\centering
	\includegraphics[width=0.45\textwidth]{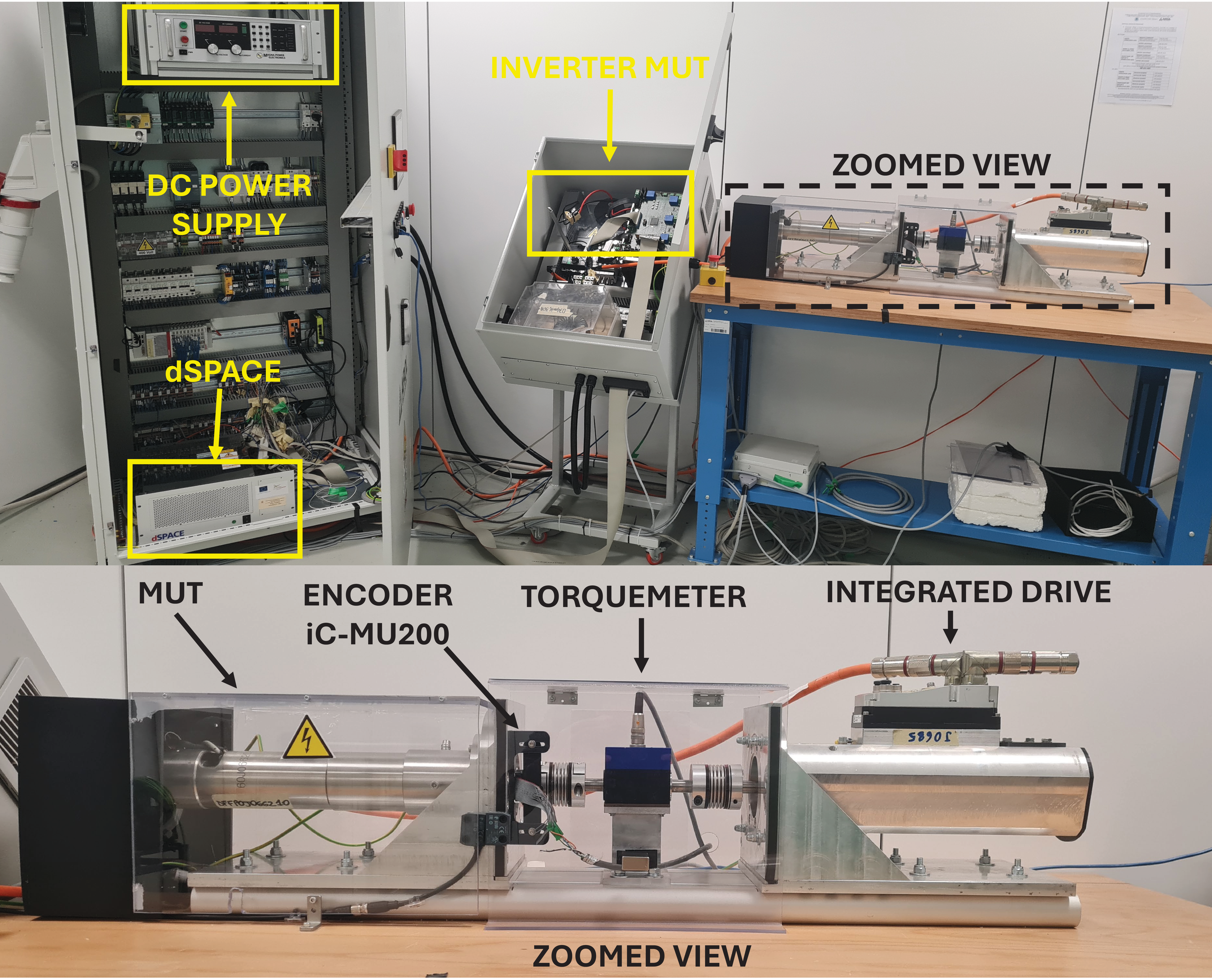}
	\caption{Experimental setup (laboratory view).}
	\label{Fig.X} 
    \end{figure}

\begin{figure}[h!]
	\centering \includegraphics[width=0.9\columnwidth]{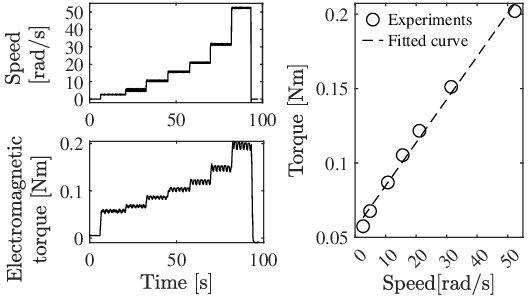}
	\caption{No-load test for friction identification: (a)-(b) measured rotor speed and electromagnetic torque during the test, (c) data fitting results.}
	\label{Fig: friction identification} 
\end{figure}

The experimental study of the load torque estimation is performed by using the setup shown in \Cref{Fig.X}. The motor under test (MUT) is a surface-mounted PMSM whose relevant parameters are listed in \Cref{tab:machine_param}. The MUT is speed-controlled by a field programmable gate array (FPGA)-based control unit (dSPACE 1006) and fed by an insulated-gate bipolar transistor-based VSI. 
The position feedback is provided by an iC-MU200 off-axis magnetic encoder.  
\begin{table}[h!]
\renewcommand{\arraystretch}{1.1}
    \centering
    \caption{Parameters of the MUT \label{tab:machine_param}}
    \begin{tabulary}{\linewidth}{C|C|C}
    \hline
    \textbf{Parameter} & 
    \textbf{Value} & 
    \textbf{Unit}\\
    \hline
    Number of pole pairs & 7        & - \\   
    DC-link voltage & 72& V \\
    Control loop frequency     & 8       & kHz \\ 
    $K_m$     & 0.6017      & Nm/A\\ 
    $J$           & $2.7354 \cdot10^{-4}$              & kgm$^2$ \\
    $b$           &$2.903\cdot10^{-3}$             & Nms/rad \\
    $C_f$  & $5.5519 \cdot10^{-2}$ & Nm \\ 
    \hline
    \end{tabulary}
\end{table}
During the experiments, the MUT is operated under different speed reference profiles, including constant, square, and sinusoidal waveforms. The load motor, equipped with an integrated drive, is controlled through its dedicated software to generate torque profiles such as constant, square, sinusoidal, and combined square–sinusoidal waveforms. The load torque is measured by using a HBK T210 shaft-mounted torque transducer which allowed to evaluate the performance of the investigated torque observers.
The torque constant $K_m$ was determined through an offline estimation. For surface-mounted PMSMs, the torque constant is given by $K_m=1.5n_{p}\psi$, where $n_p$ is the number of pole pairs and $\psi$ is the rotor flux linkage. Hence, to determine $K_m$, the rotor flux linkage was estimated by means of an open-circuit speed test. The coefficients $b$ and $C_f$ of the MUT were determined through a no-load test performed by disconnecting the motor from both the torquemeter and the load motor. Under these conditions, the following steady-state equation holds
\begin{equation}
K_m i_q = b \dot{q} + C_f\mathrm{sign}(\dot{q}).
\end{equation}
By carrying out the test at different speeds and using the measured values of $i_q$ together with the known estimate of $K_m$, the coefficients $b$ and $C_f$ were obtained through a least-squares approach. Figure \ref{Fig: friction identification} illustrates the experimental waveforms of the speed and electromagnetic torque during the test and the observation points in torque–speed plane along with the fitted curve, calculated for the determined values of $b$ and $C_f$. As can be seen, the fitted curve is in good agreement with the trend exhibited by the experimental data.
\color{black}Finally, the rotor inertia $J$ was identified through a constant electromagnetic torque acceleration test, assuming a constant rotor acceleration over a short time interval, approximated by using finite differences, and exploiting the previously estimated parameters $K_m$, $b$, $C_f$.

To eliminate the undesired noise in the measurements, digital low-pass filters were implemented to process the motor current and load torque signals. Specifically, a fast first-order filter with a time constant of 0.0025 sec was used for the motor current, while a moving average filter with a window length of 50 samples was applied to the load torque.\color{black}

\section{Linear observation techniques}  
\label{sec:3}

\subsection{Reduced-order Luenberger observer}  
\label{sec:3:sub:1}

The reduced-order Luenberger observer, see \cite{luenberger1971} for details, assumes a state-space system model in the regular form
\begin{equation}
\left(%
\begin{array}{c}
  \dot{\bar{w}} \\
  \dot{y} \\
\end{array}%
\right) =
\left(%
\begin{array}{cc}
A_{11}     & A_{12}  \\
A_{21}     & A_{22}
\end{array}%
\right) \left(%
\begin{array}{c}
  \bar{w} \\
  y \\
\end{array}%
\right) + \left(%
\begin{array}{c}
  B_{\bar{w}} \\
  B_y \\
\end{array}%
\right) \, u.
 \label{eq:3:1:1}
\end{equation}
Here, $\bar{w}$ is the vector of measurable states, i.e. those which sensor values can be used and do not need to be observed, and $y$ is the vector of remaining system states which values need to be observed. This allows an exclusion of $\bar{w}$ from the states estimate and, this way, reduces the observer order from $n$ to $m < n$, when $\dim(y) = m$ and $\dim(\bar{w}) + \dim(y) = n$. The reduced-order Luenberger observer is then given by \cite{luenberger1971}:
\begin{eqnarray}
\label{eq:3:1:2}
\dot{\breve{y}} &=& (A_{22} - K A_{12}) \breve{y} + (B_y - K B_{\bar{w}}) u + \\
\nonumber   & & (A_{21} - K A_{11} + A_{22} K - K A_{12} K) \bar{w},
\end{eqnarray}
where $K \in \mathbb{R}^{m}$ is the vector of the observer gains to be assigned. The dynamic variable $\breve{y}(t)$ is the estimate vector of the unmeasurable system states. Since $\breve{y}$ was transformed according to the right-hand-side of \eqref{eq:3:1:2}, a back transformation
\begin{equation}\label{eq:3:1:3}
\tilde{y}(t) = \breve{y}(t) + K \bar{w}(t)
\end{equation}
is also required, in order to obtain the observed system states $\tilde{y}(t) \rightarrow y(t)$ of interest. Obviously, the natural dynamics of the reduced-order Luenberger observer and, therefore, its stable asymptotic convergence, are determined by designing the observer system matrix $(A_{22} - K A_{12})$ to be Hurwitz, while two other bracket-terms on the right-hand-side of \eqref{eq:3:1:2} are driven by the exogenous quantities $u$ and $\bar{w}$.

Using a well-established approach \cite{meditch1974observers} for the unknown time-varying
disturbances, following to which a zero-order dynamics is assumed for the load torque state, i.e. $\dot{\tau}=0$, the linear part of the system \eqref{eq:2:1} can be transformed into the regular normal form \eqref{eq:3:1:1}  with $\bar{w} = q$ and $y = [\dot{q}, \tau]^\top$ as
\begin{equation}
\left(%
\begin{array}{c}
  \dot{q} \\[1mm]
  \ddot{q} \\[1mm]
  \dot{\tau}\\[1mm]
\end{array}%
\right) =
\left(%
\begin{array}{ccc}
0  & 0 & 1  \\[1.5mm]
0  & -\dfrac{b}{J} & \dfrac{1}{J} \\[3mm]
0  & 0 & 0\\[1.5mm]
\end{array}%
\right) \left(%
\begin{array}{c}
  q \\[1mm]
  \dot{q} \\[1mm]
  \tau \\[1mm]
\end{array}%
\right) + \left(%
\begin{array}{c}
  0 \\[1.5mm]
  \dfrac{1}{J} \\[3mm]
  0 \\[1.5mm]
\end{array}%
\right) \, \bar{u}.
 \label{eq:3:1:4}
\end{equation}
Note that the used input $\bar{u}$ excludes the (modeled) nonlinear friction term, cf. \eqref{eq:2:1}, which is matched and, hence, can be subtracted prior to the observer, cf. below with Fig. \ref{fig:1}. Then, the designed reduced-order Luenberger observer \eqref{eq:3:1:2}, \eqref{eq:3:1:3} with the gain vector $K = [K_1, K_2]^\top$ results in
\begin{align}
\label{eq:3:1:5}
\dot{\breve{y}}  = &
\left(%
\begin{array}{cc}
-K_1-\dfrac{b}{J}  &  \dfrac{1}{J}  \\[4mm]
-K_2               &  0
\end{array}%
\right) \breve{y} + \left(%
\begin{array}{c}
  \dfrac{1}{J} \\[4mm]
  0
\end{array}%
\right) \bar{u} + M  q \quad \hbox{ with}  \\[2mm]
\nonumber M = & \left[%
\begin{array}{c}
  -K_1^2 - \dfrac{b K_1 - K_2}{J} \\[4mm]
  -K_1 K_2
\end{array}%
\right], \quad 
\Bigl[\tilde{\dot{q} },\, \tilde{\tau}\Bigr]^\top =  \breve{y} + \left[%
\begin{array}{c}
  K_1 \\[1mm]
  K_2
\end{array}%
\right] q.
\end{align}
The observer dynamics can directly be shaped by any conventional linear feedback technique, i.e. either by poles assignment of the observer system matrix, or by LQR optimization, or by also customizing any particular (e.g. application-specific) criteria of the transient response and convergence of the state estimates $\tilde{\dot{q}}(t)$ and $\tilde{\tau}(t)$ towards the actual values.

\subsection{Disturbance observer}  
\label{sec:3:sub:3}

The disturbance observer \cite{ohishi1987} (often abbreviated as DOB) assumes a stable input-output transfer function
\begin{equation}\label{eq:3:3:1}
\Sigma(s) = q(s) \, u(s)^{-1} = C \bigl(s I  - A \bigr)^{-1} B
\end{equation}
of the modeled plant, where a standard state-space notation of the $n$-th order system is given by $A \in \mathbb{R}^{n \times n}$, $B \in \mathbb{R}^{n \times 1}$, $C \in \mathbb{R}^{1 \times n}$. $I$ is the identity matrix and the argument $s$ is the complex Laplace variable. It is required that the system \eqref{eq:3:3:1}, including also the case of an uncertain plant, is minimum phase, cf. \cite{shim2009}, so as to have $\Sigma(s)$ invertible stable.   

If the linear system input $u(t)$ is additively affected by a matched but unknown (i.e. not measurable) disturbance $\tau(t)$, cf. also the considered actuator case \eqref{eq:2:1}, then DOB aims to reconstruct the disturbance value, respectively compute its estimate $\tilde{\tau}(t)$. The latter is mapped, through the block diagram algebra, by the difference between the available input and back-propagation of the measured output through the inverse system model, i.e.
\begin{equation}\label{eq:3:3:2}
\bigl(u(s) - \Sigma^{-1}(s) q(s)\bigr) \mapsto \tilde{\tau}(s).
\end{equation}
An inherent problem is that $\Sigma^{-1}(s)$ can be improper, that is usually the case when a physical plant has a relative degree (see e.g. \cite{doyle2009} for details) higher than zero. Therefore, a low-pass filter $Q(s)$ is always required by the DOB, so that the transfer function $Q(s) \Sigma^{-1}$ becomes proper and implementable \cite{shim2009}. Note that the order of $Q(s)$ must be equal or higher than the relative degree of $\Sigma(s)$, and $Q(0) = 1$. Since the  introduction of DOB \cite{ohishi1987} and its further popularisation \cite{ohnishi1994}, an appropriate design of the $Q$-filter became a pivot point of multiple related works (not listed here for the space reasons, see e.g. several references provided in \cite{ruderman2020}). With use of the low-pass filter, the DOB transfer characteristics results is 
\begin{equation}\label{eq:3:3:3}
\tilde{\tau}(s) = Q(s) u(s) - Q(s) \Sigma^{-1}(s) q(s).
\end{equation}

Considering the linear part of the system \eqref{eq:2:1}, and assuming the most straightforward second-order low-pass filter $Q(s)$, with the critical damping and natural frequency $\omega_0$ as the design parameter, the synthesized DOB cf. \eqref{eq:3:3:3} yields 
\begin{equation}\label{eq:3:3:6}
\tilde{\tau}(s) = \dfrac{\omega_0^2}{s^2 + 2 \omega_0 s + \omega_0^2} \bar{u}(s) - \dfrac{\omega_0^2 \, \bigl(J s^2 + b s \bigr) }{K_m \bigl(s^2 + 2 \omega_0 s + \omega_0^2\bigr)} q(s).
\end{equation}
Note that $\bar{u}$ is used instead of $u$, similar as before in Luenberger observers cf. Fig. \ref{fig:1}, and the implementation of \eqref{eq:3:3:6} enables also for certain flexibility in terms of the sequence of the involved integration and differentiation operations. Thus, the $Q(s) s\, q(s)$-step can be performed first, correspondingly separately, in order to obtain the estimate of relative velocity $\tilde{\dot{q}}$, see below in section \ref{sec:4}.

\section{Observer in the loop}  
\label{sec:4}

\begin{figure}[!h]
\centering
\includegraphics[width=0.9\columnwidth]{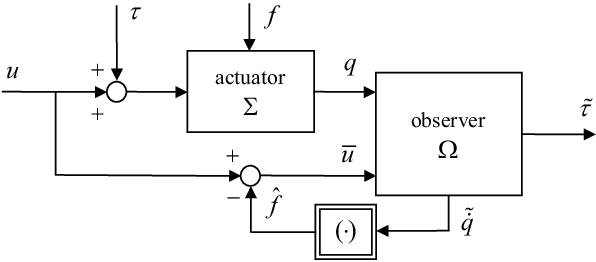}
\caption{Block diagram of the load torque observer in an open-loop.}
\label{fig:1}
\end{figure}

Without loss of generality, any type of the input-output observer (linear in our case) can be seen as a map $\Omega: \: \bigl(\bar{u}(t),q(t)\bigr) \mapsto \tilde{\tau}(t)$, where the available input of a linear system is $\bar{u}$ and the measured output system state is $q$. The observer output is then the estimated quantity $\tilde{\tau}$, of the load torque in our case. Denoting the actuator system by $\Sigma: \: \bigl(u(t)+\tau(t)\bigr) \mapsto q(t)$, cf. \eqref{eq:2:1}, the load torque observer in open-loop results in the block diagram represented in Fig. \ref{fig:1}.  
Note that since the linear observation techniques are in our primary focus, the nonlinear friction torque $f(\cdot)$ is regarded as an additional (matched) disturbance value, which model is however available. Therefore, the computed $\hat{f}(\cdot)$ friction torque is subtracted from the input channel of any of the used observers $\Omega$ leading to $\bar{u}= u - \hat{f}$, provided the estimated relative velocity $\tilde{\dot{q}}$ is available. Further we note that while the most simple (yet general and reliable enough) Coulomb friction model is used, cf. \eqref{eq:2:1}, the observer structure as in Fig. \ref{fig:1} allows also for a more detailed nonlinear friction modeling, e.g., including the pre-sliding transitions, cf. \cite{ruderman2023analysis}. 

Since one of the main objectives of a load torque observer is to use the estimated torque value $\tilde{\tau}(t)$ for a subsequent compensation, i.e. by applying $\tilde{\tau}$ to the control channel $\bar{u}$, it is natural to look into the open-loop transfer characteristics 
\begin{equation}\label{eq:4:1}
L(s) = \tilde{\tau}(s) \, \bar{u}(s)^{-1}
\end{equation}
of each particular observer $\Omega$. Such approach was also pursued in \cite{ruderman2024}, while purposefully revealing a practical unsuitability of an observer-based feedback control despite the system itself proved to be fully observable. The feed-forward signal path can exclude entirely the injection of nonlinear friction, cf. Fig. \ref{fig:1}, so that a transfer function $L(s)$ can be derived for the linear actuator's dynamics $\Sigma$ and the respective observer $\Omega$. Below, $L_{\{L,D\}}(s)$ characteristics are determined for both observation techniques under consideration, while the sub-indices $\{L,D\}$ denote the Luenberger and disturbance observers, cf. Sections \ref{sec:3:sub:1} and \ref{sec:3:sub:3}, respectively. Especially, the impact of (unavoidable) uncertainties are shown for both observers $\Omega_{\{L,D\}}$.

\subsection{Stability margins of observer open-loop}  
\label{sec:4:sub:1}

First, one needs to stress that any real actuator system can have parametric uncertainties $\Delta_{K_m}$, $\Delta_{J}$, and $\Delta_b$, cf. \eqref{eq:2:1}, while a physically reasonable assumption will constrain 
\begin{equation}\label{eq:4:0:1}
K_m - \bigl|  \Delta_{K_m} \bigr| > 0, \quad J - \bigl|  \Delta_{J} \bigr| > 0, \quad b - \bigl|  \Delta_{b} \bigr| > 0.
\end{equation}
Following to that, the transfer function 
\begin{equation}\label{eq:4:0:2}
\Sigma(s) = \dfrac{K_m + \Delta_{K_m}}{\bigl(J + \Delta_{J}\bigr) s^2  + \bigl(b + \Delta_{b}\bigr) s}
\end{equation}
appears in \eqref{eq:3:3:1} for describing the effective system plant.

As a more sensitive case, we consider next the loop transfer function \eqref{eq:4:1} for $\Omega_D$, which results in 
\begin{equation}\label{eq:4:3:1}
L_{D}(s) = \Bigl(1 - \Sigma(s) \tilde{\Sigma}^{-1}(s)\Bigr) Q(s)
\end{equation}
and is equal to zero if and only if $\Sigma = \tilde{\Sigma}$. Recall that $\tilde{\Sigma}$ is the nominal (i.e. initially identified) system transfer function and an uncertain plant $\Sigma \neq \tilde{\Sigma}$ will always be appearing, so that the loop transfer function \eqref{eq:4:3:1} leads to  
\begin{equation}\label{eq:4:3:2}
L_{D}(s) =  \dfrac{\bigl(K_m \Delta_J - J \Delta_{K_m} \bigr) s + \bigl(K_m \Delta_b - b \Delta_{K_m}\bigr)}{\bigl(J + \Delta_J\bigr) s + \bigl(b + \Delta_b\bigr)} \,\dfrac{Q(s)}{K_m}.
\end{equation} 
From the parametric assumption \eqref{eq:4:0:1}, one can recognize that the pole of \eqref{eq:4:3:2} remains always stable while, at the same time, $L_{D}(s)$ can easily turn to have an unstable zero, thus becoming non-minimum phase. Since the low-pass filter $Q(s)$ must be a stable minimum-phase transfer function, the single yet essential criteria for $L_{D}(s)$ remains minimum-phase is
\begin{equation}\label{eq:4:3:3}
\dfrac{K_m \Delta_b - b \Delta_{K_m}}{K_m \Delta_J - J \Delta_{K_m}} > 0.
\end{equation}
Since the implication of unstable zeros (and so non-minimum phase behavior) on the stability and performance of a closed-loop system are well known in the control theory, a further detailed analysis of violating the condition \eqref{eq:4:3:3} is omitted here. Solely recall that a right half-plane zero gives an upper bound to the achievable bandwidth, and the bandwidth decreases with decreasing frequency of the unstable zero \cite{aastrom2000}. 

For the sake for illustration, however, we show how sensitive the $L_{D}(s)$ can be already the case if only two parameters of the plant model are uncertain, and one of which only minor. For the sake of simplicity, the total inertia is kept certain, i.e. $\Delta_{J} = 0$, that can be a realistic assumption for any designed actuator without load. The motor torque constant is assumed to have $\Delta_{K_m} = \pm 0.1 K_m$, while the damping coefficient is assumed being once increasing and once decreasing by 1/3 of the nominal value, i.e. $\Delta_{b} = \pm b/3$. Note that the latter is more than realistic given the fact of usually time-varying and highly uncertain viscous damping conditions in the actuator systems. The experimentally identified (nominal) parameter values are assumed for $K_m$, $J$, $b$, see section \ref{sec:2:sub:2}, while the dynamics of the $Q(s)$-filter is assigned to have the double real pole at $-500$, i.e. almost the same as the poles of the designed reduced-order Luenberger observer, cf. section \ref{sec:5:sub:1}. A maximal (critical) gaining factor          
$\max k_c$ of the loop transfer function $L_{D}(s)$, beyond which the closed-loop system becomes unstable, is evaluated as a stability margin. Note that $k_c$ can capture additional $K_m$ uncertainties but not only those, cf. \eqref{eq:4:3:2}. One can also recognize that the cut-off frequency of $Q(s)$, which has the unity gain, has only minor effect on the stability margin of $L_{D}(s)$. The evaluated $\max k_c$ values are listed in Table \ref{tab:2} in dependency of the parametric uncertainties.
\begin{table}[h!]
\renewcommand{\arraystretch}{1.2}
    \centering
    \caption{}
    \begin{tabulary}{\linewidth}{|C||C|C|C|C|}
    \hline
    $\Delta_{K_m}$ & $+ 1 \%$ & $+ 1 \%$ & $- 1 \%$ & $- 1 \%$ \\
    \hline 
    $\Delta_{b}$ &  $+ 33 \%$ &   $- 33 \%$ &  $+ 33 \%$ &  $-33 \%$ \\
    \hline \hline
    $\max k_c$ & 10 & 1.6 & inf & 1.6 \\
    \hline
    \end{tabulary}
    \label{tab:2}
\end{table}
Note that the assigned damping coefficient either can turn $L_{D}(s)$ to be non-minimum phase, i.e. \eqref{eq:4:3:3} becomes violated, while the frequency of the unstable zero is relatively low, cf. \cite{aastrom2000}. Or, it can drastically reduce the gain margin. The situation can further worsen, in terms of stability margins, if additional (even minor) time delays or parasitic dynamics (e.g. due electro-magnetic motor circuits, output sensor, and digital observer implementation) appear in the loop transfer function.  

For the reduced-order Luenberger observer $\Omega_L$, the loop transfer function \eqref{eq:4:1} yields (cf. Fig. \ref{fig:1} and \cite[section~III.B]{ruderman2024})
\begin{equation}\label{eq:4:1:1}
L_{L}(s) = \bigl[ 0 \;  1 \bigr]
\bigl(s I  - A_L  \bigr)^{-1} \Bigl[ B_L \:  M \Bigr]
\left(
             \begin{array}{c}
               1 \\
               \Sigma \\
             \end{array}
           \right),
\end{equation}
where the observer system matrix $A_L$ and the input coupling vector $B_L$ are those from the first equation in \eqref{eq:3:1:5}. Upon reducing the overall composed loop transfer function \eqref{eq:4:1:1}, one can shown that it has two stable poles (including one in the origin), one very  fast unstable zero, and one even faster stable zero. Remarkable is the fact that such pole-zero configuration of $L_L(s)$ changes very little if the above parametric uncertainties in 
\eqref{eq:4:0:2} are in place. Since the frequency of unstable zero is relatively high, the bandwidth is not significantly reduced, cf. \cite{aastrom2000}, in contrast to the non-minimum phase issues of the DOB, as shown above. The same illustrative parametric variations $\Delta_{K_m} = \pm 0.1 K_m$, $\Delta_{b} = \pm b/3$ as above are evaluated for $L_L(s)$, while also the same maximal (critical) gaining factor $\max k_c$ is considered as stability margin. The results are listed in Table \ref{tab:3} in dependency of the parametric uncertainties.
\begin{table}[h!]
\renewcommand{\arraystretch}{1.2}
    \centering
    \caption{}
    \begin{tabulary}{\linewidth}{|C||C|C|C|C|}
    \hline
    $\Delta_{K_m}$ & $+ 1 \%$ & $+ 1 \%$ & $- 1 \%$ & $- 1 \%$ \\
    \hline 
    $\Delta_{b}$ &  $+ 33 \%$ &   $- 33 \%$ &  $+ 33 \%$ &  $-33 \%$ \\
    \hline \hline
    $\max k_c$ & inf & inf & inf & inf \\
    \hline
    $\phi$ (deg) & 97.3 & 96.9 & 95.7 & 95.4 \\
    \hline
    \end{tabulary}
    \label{tab:3}
\end{table}

It must be fairly noted that the phase margin $\phi = 180 + \angle L(j\omega_c)$ must also to be taken into account for $L_L(s)$, while for $L_D(s)$ it is not. Indeed, $|L_D(s)| < 1$ for all frequencies, independent of the parametric variations, so that its phase margin is theoretically infinite. Quite the opposite, $|L_L(s)|$ always experiences a cross-over frequency $\omega_c$ so that $\phi$ needs to be additionally evaluated. However, $\phi$ turns to be always sufficiently high for $L_L(s)$, i.e. $> 90$ deg, cf. Table \ref{tab:3}.

\section{Comparative results}  
\label{sec:5}

\subsection{Assigned observer parameters}  
\label{sec:5:sub:1}

Both designed observers are parameterized in the sense to have a comparable transient behavior  and so the convergence of the load torque estimate $\tilde{\tau}(t)$. For the reduced-order Luenberger observer \eqref{eq:3:1:5}, both poles of the observer system matrix are placed at $\lambda_{1,2} = 50 \times \min\{ p_i \}$, where $p_i$ are the poles of the identified linear system. This way, the observer poles are, to say, fifty times faster than the fastest pole of the system to be observed. For the identified system model $\min\{ p_i \} = -10.61$. Also both poles of the $Q$-filter in DOB, and hence its natural frequency, are placed at $-500$, i.e.  $\omega_0 = 500$ rad/s, cf. \eqref{eq:3:3:6}.

\subsection{Experimental observers evaluation}  
\label{sec:5:sub:2}

The experimental evaluation is running for the controlled speed about 10.5 rad/sec, while a dynamic variation of the load torque is realized via the load motor, cf. section \ref{sec:2:sub:2}. Note that $\tau(t)$ is measured (for evaluation) by means of the torquemeter, see Fig. \ref{Fig.X}, thus representing effectively the load torque acting on the shaft of the actuator, i.e. MUT.      

\begin{figure}[!h]
\centering
\includegraphics[width=0.98\columnwidth]{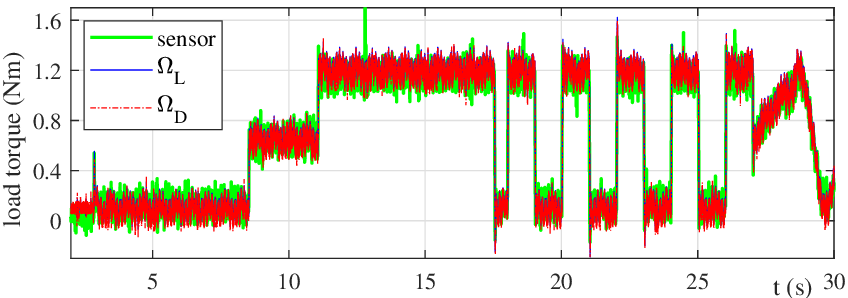}
\includegraphics[width=0.98\columnwidth]{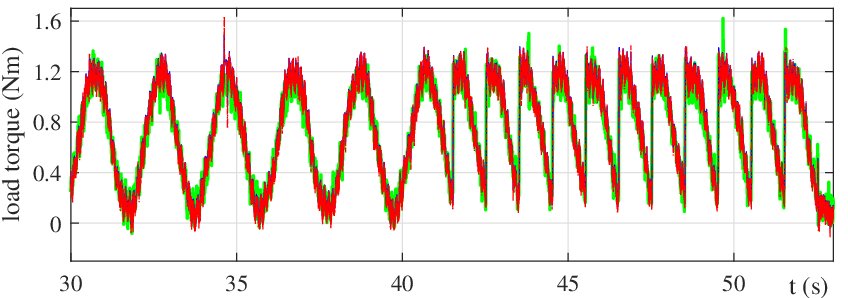}
\caption{Measured and estimated (with $\Omega_L,\, \Omega_D$ observers) load torque from the experiments with the controlled speed about 10.5 rad/sec.}
\label{fig:TorqueExp}
\end{figure}

The measured and estimated load torque are depicted over each other in Fig. \ref{fig:TorqueExp} for both cases, of the reduced-order Luenberger observer $\Omega_L$ and DOB $\Omega_D$.
Note that since the long-term patterns are visually very similar for both observers, a more detailed visualization at steady-state phases will be omitted for the sake of brevity. Instead, the striking cutouts of some transient phases will be compared graphically below, in addition to the evaluated root mean square (RMS) error
$$
\textrm{RMS} = \sqrt{ \dfrac{1}{T} \int_{0}^{T} \Bigl(\tau(t)-\tilde{\tau}(t)\Bigr)^2 dt }.
$$
For a quantitative measure of the estimation error spread, the standard deviation (STD) of the torque observer absolute error $\bigl|\tau(t)-\tilde{\tau}(t)\bigr|$ is also evaluated. Both estimation error metrics are summarized in Table \ref{tab:4}.
\begin{table}[h!]
\renewcommand{\arraystretch}{1.2}
    \centering
    \caption{Load torque estimation error metrics}
    \begin{tabulary}{\linewidth}{|C||C|C|}
    \hline
    Observer  & $\Omega_L$    & $\Omega_D$  \\
    \hline \hline
    RMS       & 0.2109        &  0.2991 \\
    \hline 
    STD       & 0.2097        &  0.2887 \\
    \hline
    \end{tabulary}
    \label{tab:4}
\end{table}

\begin{figure}[!h]
\centering
\includegraphics[width=0.98\columnwidth]{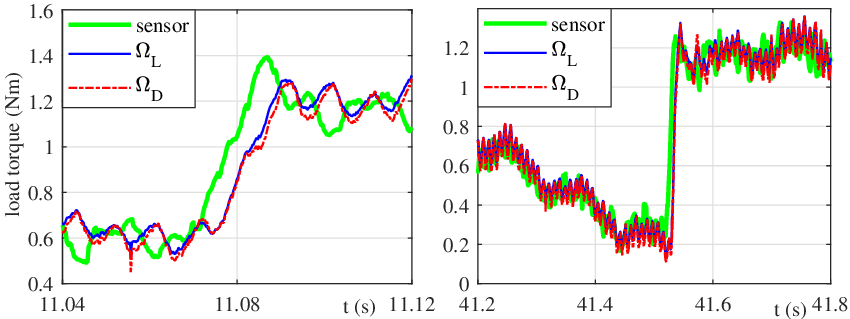}
\caption{Zoom-in of the measured and estimated (with $\Omega_{\{L,D\}}$ observers) load torque from the experiments as depicted in Fig. \ref{fig:TorqueExp}.}
\label{fig:TorqueZoomed}
\end{figure}

Two illustrative zooms-ins of the experiments, the same as shown in Fig. \ref{fig:TorqueExp}, are additionally depicted in Fig. \ref{fig:TorqueZoomed} for both observers $\Omega_{\{L,D\}}$. On the left, one can recognize a transient phase, correspondingly time lag of $< 0.01$ sec for both observers. It is also visible that the fast sub-harmonics are followed, but being mostly in the anti-phase. On the right, it is visible that both observers are following closely the slower harmonics of the load torque, cf. also with Fig. \ref{fig:TorqueExp}.  

\section{Conclusions}  
\label{sec:6}

In this paper, we considered two linear observation techniques, reduced-order Luenberger observer \cite{luenberger1971} and disturbance observer \cite{ohishi1987}, and made a comparison when using them for robust estimation of the load torque in rotary actuator systems. To this end, both observation methods were addressed from the observer loop transfer function viewpoint, while the related aspects of eventually non-minimum phase behavior and stability margins were highlighted. The analysis disclosed that DOB is more sensitive to the parametric uncertainties of the nominal (i.e. identified) system plant, in terms of the gain margin reduction. Contrary, the reduced-order Luenberger observer has infinity gain margin in face of the parametric uncertainties, while the phase margin needs also to be taken into account. The latter remains however $> 90$ deg irrespective uncertainties in the shown case study. For a fair comparison, both observers were designed to have the same number of the real poles placed in the same location. The detailed experimental case study, accomplished on a PMSM-based actuator drive with a counteracting active load and torquemeter for reference measurements, confirms our analysis and shows a comparative observers' evaluation in a detailed and transparent manner.

Future works can be concerned with an extended comparison with some further, hybrid (i.e. linear-nonlinear) methods, e.g. detection and isolation \cite{deluca2003,ruderman2014}, and moreover with an explicit sensitivity analysis in view of both, parametric system uncertainties and the measurement noise. The overall objective is to employ and compare such observations techniques in a feedback compensation of the unknown load torques.


\bibliographystyle{IEEEtran}
\bibliography{references}

\begin{thebibliography}{10}
\providecommand{\url}[1]{#1}
\csname url@rmstyle\endcsname
\providecommand{\newblock}{\relax}
\providecommand{\bibinfo}[2]{#2}
\providecommand\BIBentrySTDinterwordspacing{\spaceskip=0pt\relax}
\providecommand\BIBentryALTinterwordstretchfactor{4}
\providecommand\BIBentryALTinterwordspacing{\spaceskip=\fontdimen2\font plus
\BIBentryALTinterwordstretchfactor\fontdimen3\font minus
  \fontdimen4\font\relax}
\providecommand\BIBforeignlanguage[2]{{%
\expandafter\ifx\csname l@#1\endcsname\relax
\typeout{** WARNING: IEEEtran.bst: No hyphenation pattern has been}%
\typeout{** loaded for the language `#1'. Using the pattern for}%
\typeout{** the default language instead.}%
\else
\language=\csname l@#1\endcsname
\fi
#2}}

\bibitem{ruderman2020}
M.~Ruderman, M.~Iwasaki, and W.-H. Chen, ``Motion-control techniques of today
  and tomorrow: a review and discussion of the challenges of controlled
  motion,'' \emph{IEEE Industrial Electronics Magazine}, vol.~14, no.~1, pp.
  41--55, 2020.

\bibitem{ohnishi1994}
K.~Ohnishi, N.~Matsui, and Y.~Hori, ``Estimation, identification, and
  sensorless control in motion control system,'' \emph{Proceedings of the
  IEEE}, vol.~82, no.~8, pp. 1253--1265, 1994.

\bibitem{luenberger1971}
D.~Luenberger, ``An introduction to observers,'' \emph{IEEE Transactions on
  Automatic Control}, vol.~16, no.~6, pp. 596--602, 1971.

\bibitem{bernard2018}
P.~Bernard and V.~Andrieu, ``Luenberger observers for nonautonomous nonlinear
  systems,'' \emph{IEEE Transactions on Automatic Control}, vol.~64, no.~1, pp.
  270--281, 2018.

\bibitem{ohishi1987}
K.~Ohishi, M.~Nakao, K.~Ohnishi, and K.~Miyachi, ``Microprocessor-controlled
  {DC} motor for load-insensitive position servo system,'' \emph{IEEE
  Transactions on Industrial Electronics}, vol.~34, no.~1, pp. 44--49, 1987.

\bibitem{oboe2018}
R.~Oboe, ``How disturbance observer changed my life,'' in \emph{IEEE 15th
  International Workshop on Advanced Motion Control}, 2018, pp. 13--20.

\bibitem{shim2009}
H.~Shim and N.~H. Jo, ``An almost necessary and sufficient condition for robust
  stability of closed-loop systems with disturbance observer,''
  \emph{Automatica}, vol.~45, no.~1, pp. 296--299, 2009.

\bibitem{depersis2002}
C.~De~Persis and A.~Isidori, ``A geometric approach to nonlinear fault
  detection and isolation,'' \emph{IEEE Transactions on Automatic Control},
  vol.~46, no.~6, pp. 853--865, 2002.

\bibitem{deluca2003}
A.~De~Luca and R.~Mattone, ``Actuator failure detection and isolation using
  generalized momenta,'' in \emph{IEEE International Conference on Robotics and
  Automation}, vol.~1, 2003, pp. 634--639.

\bibitem{ruderman2014}
M.~Ruderman, T.~Bertram, and M.~Iwasaki, ``Modeling, observation, and control
  of hysteresis torsion in elastic robot joints,'' \emph{Mechatronics},
  vol.~24, no.~5, pp. 407--415, 2014.

\bibitem{ruderman2023analysis}
M.~Ruderman, \emph{Analysis and compensation of kinetic friction in robotic and
  mechatronic control systems}.\hskip 1em plus 0.5em minus 0.4em\relax CRC
  Press, 2023.

\bibitem{ruderman2025}
M.~Ruderman, ``On convergence analysis of feedback control with integral action
  and discontinuous relay perturbation,'' \emph{Commun. in Nonlin, Science and
  Numerical Simulation}, vol. 145, p. 108698, 2025.

\bibitem{meditch1974observers}
J.~Meditch and G.~Hostetter, ``Observers for systems with unknown and
  inaccessible inputs,'' \emph{International Journal of Control}, vol.~19,
  no.~3, pp. 473--480, 1974.

\bibitem{doyle2009}
J.~Doyle, B.~Francis, and A.~Tannenbaum, \emph{Feedback Control Theory}.\hskip
  1em plus 0.5em minus 0.4em\relax Dover Publications, 2009.

\bibitem{ruderman2024}
M.~Ruderman, ``Adaptive time delay based control of non-collocated oscillatory
  systems,'' in \emph{IEEE Mediterranean Conference on Control and Automation
  (MED2024)}, 2024, pp. 125--130.

\bibitem{aastrom2000}
K.~J. {\AA}str{\"o}m, ``Limitations on control system performance,''
  \emph{European Journal of Control}, vol.~6, no.~1, pp. 2--20, 2000.

\end{thebibliography}

\end{document}